\documentclass[conference]{IEEEtran}
\IEEEoverridecommandlockouts
\usepackage{array}
\usepackage{makecell}
\usepackage{xcolor}
\usepackage{colortbl}
\usepackage{hhline}
\usepackage{graphicx}
\usepackage{balance}
\usepackage{booktabs}
\usepackage{url}
\usepackage{tikz}

\usepackage[most]{tcolorbox}

\definecolor{mygreen}{HTML}{CCFFCC}
\definecolor{mygreen2}{HTML}{EEFFEE}

\definecolor{aurometalsaurus}{rgb}{0.43, 0.5, 0.5}
\definecolor{lightgray}{rgb}{0.94, 0.94, 0.94}

\newtcolorbox{mybox}[1]{enhanced,attach boxed title to top center={xshift=-28mm,yshift=-3mm,yshifttext=-1mm},
width=0.98\linewidth,
  colback=lightgray,colframe=black,colbacktitle=aurometalsaurus,
  fonttitle=\bfseries,
  title=#1,
  boxed title style={size=small,colframe=black},
  before=\begin{center},after=\end{center}
 }
 
\newcommand\copyrighttextbottom{%
  \footnotesize Accepted version of R. Micko, S. Chren and B. Rossi, "Applicability of Software Reliability Growth Models to Open Source Software," to appear in 48th Euromicro Conference Series on Software Engineering and Advanced Applications (SEAA'22)}
\newcommand\copyrightnoticebottom{%
\begin{tikzpicture}[remember picture,overlay]
\node[anchor=south,yshift=2pt] at (current page.south) {\fbox{\parbox{\dimexpr1.0\textwidth-\fboxsep-\fboxrule\relax}{\copyrighttextbottom}}};
\end{tikzpicture}%
}

\begin{document}

\title{Applicability of Software Reliability Growth Models to Open Source Software}

\author{\IEEEauthorblockN{Radoslav Mi\v{c}ko}
\IEEEauthorblockA{\textit{Faculty of Informatics} \\
\textit{Masaryk University}\\
Brno, Czechia \\
r.micko@mail.muni.cz}
\and
\IEEEauthorblockN{Stanislav Chren}
\IEEEauthorblockA{\textit{Faculty of Informatics} \\
\textit{Masaryk University}\\
Brno, Czechia  \\
chren@mail.muni.cz}
\and
\IEEEauthorblockN{Bruno Rossi}
\IEEEauthorblockA{\textit{Faculty of Informatics} \\
\textit{Masaryk University}\\
Brno, Czechia  \\
brossi@mail.muni.cz}
}

\maketitle

\copyrightnoticebottom

\begin{abstract}
Software Reliability Growth Models (SRGMs) are based on underlying assumptions which make them typically more suited for quality evaluation of closed-source projects and their development lifecycles. Their usage in open-source software (OSS) projects is a subject of debate. Although the studies investigating the SRGMs applicability in OSS context do exist, they are limited by the number of models and projects considered which might lead to inconclusive results. In this paper, we present an experimental study of SRGMs applicability to a total of 88 OSS projects, comparing nine SRGMs, looking at the stability of the best models on the whole projects, on releases, on different domains, and according to different projects' attributes. With the aid of the STRAIT tool, we automated repository mining, data processing, and SRGM analysis for better reproducibility. Overall, we found good applicability of SRGMs to OSS, but with different performance when segmenting the dataset into releases and domains, highlighting the difficulty in generalizing the findings and in the search for \textit{one-fits-all} models.
\end{abstract}


\section{Introduction}
Software Reliability is one of the most important attributes of software systems. Software Reliability Growth Models (SRGMs) emerged in the '70es when Jelinski and Moranda proposed the JM model for fitting cumulative failure data aiming at providing a methodology to analyze and predict project failures over time~\cite{LYU95,jelinski}. 

However, during the years, the software development process has been evolving considering more agile and open source development processes, challenging many of the assumptions of the models that were developed over time --- like the fact that the defect fixing process does not introduce new defects or that no new code is introduced during the testing process~\cite{WOOD96}. 

The goal of this paper is to investigate the applicability of Software Reliability Growth Models (SRGM) to a large set of Open Source Software (OSS) projects. For the analysis, we adopted and customized STRAIT~\cite{chren_Strait}, a previously developed tool to provide SRGM analysis. With the aid of the tool, we fit 9 different SRGM models to cumulative failure data from 88 OSS projects. Overall, considering projects as a whole and software releases, we fit $1\,053$ SRGMs --- that, to our knowledge represents the largest empirical study on OSS SRGMs. We further provide all the datasets and outputs from the STRAIT tool~\cite{dataset2022}.
The three main contributions of this paper go into studying SRGMs in more detail in i) project domains, ii) releases, all in the context of a iii) large-scale study compared to related works.

\begin{itemize}
    \item[{i)}] Domains: we want to investigate whether different domains might follow different lifecycles and have different operational profiles thus resulting in varied issue-reporting patterns, which might affect the SRGM analysis.
    \item[{ii)}] Project releases: considering project releases can make the reliability growth trend more prominent, thus affecting the parameter estimation and the GoF. Some of the related studies conducted the SRGM analysis on the issue reports from the whole project history, some divided the data based on releases. We wanted to replicate both approaches on a larger sample of projects.
    \item[{iii)}] Large scale: in this paper, we demonstrate a large-scale SRGM analysis ($1\,053$ SRGMs, $\sim$$400K$ software defects) with the help of the STRAIT tool~\cite{chren_Strait} to motivate future replicability of the results and reuse of datasets in different SRGM studies.
\end{itemize}

The paper is structured as follows. In Section \ref{sec:background}, we provide a brief background for the SRGM process. In Section \ref{sec:sgrm-oss} we discuss the peculiarities of OSS for the applicability of SRGMs. In Section \ref{sec:related-work}, we summarize the related work to further motivate our study. Section \ref{sec:evaluation} details the methodology of the experiments with the results presented in Section \ref{sec:results}. We discuss the potential threats to validity in Section \ref{sec:threats} and conclusions in Section \ref{sec:conclusion}.

\section{Software Reliability Growth Modelling}
\label{sec:background}

During the testing and early operation phases of the software life cycle, failure events are encountered. They are recorded and the underlying faults that caused them are removed, which results in a process called reliability growth. SRGMs are regression-based models whose purpose is to estimate the parameters of a mean value function $m(t)$ based on the input data: $m(t)$ represents cumulative number of faults detected by the given time $t$. 

In this study, we consider nine models shown in Table \ref{tab:models}. These are the most important models that were defined over time to study reliability growth in cumulative failure data.

\begin{center}
\begin{table}[htb]
\caption{List of applied SRGMs}
  	\label{tab:models}
    \scriptsize
    \setlength\arrayrulewidth{1pt}
    \begin{tabular}{|c|c|c|}
        \hline
        \textbf{Model} & \textbf{Type} & \textbf{$\mu$(t)} \\
        \hline
        Goel-Okumoto (GO) \cite{GO79}& Concave & $a(1-e^{-bt})$ \\\hline
        Goel-Okumoto S-Shaped (GOS) \cite{GO79} & S-Shaped & ${a(1-}$ ${(1+bt)e^{-bt})}$ \\\hline
        Hossain-Dahiya (HD) \cite{Hoss93} & Concave & ${a(1-e^{-bt})/}$ ${(1+ce^{-bt})}$ \\\hline
        Musa-Okumoto (MO) \cite{Musa84} & Infinite & $\alpha ln(\beta t + 1)$ \\\hline
        Duane (DU) \cite{Duane64} & Infinite & $\alpha t^\beta$  \\\hline
        Weibull (WE) \cite{musa87} & Concave & $a(1-e^{-bt^c})$ \\\hline
        Yamada Exponential (YE) \cite{yamada83} & Concave & ${a(1-e^{-r\alpha (1 - e ^{-\beta t})})}$ \\\hline
        Yamada Raleigh (YR) \cite{yamada83} & S-Shaped & ${a(1-e^{-r\alpha (1 - e ^{-\beta t^2 /2})})}$ \\\hline
        Log-Logistic (LL) \cite{log98} & S-Shaped & ${a(\lambda t)^\kappa /}$ ${(1+(\lambda t)^\kappa )}$ \\
        \hline
    \end{tabular}
    
\end{table}
\end{center}

Based on the shape of $m(t)$, we can classify SRGMs into three main categories \cite{LYU95} 
\begin{itemize}
\item \textit{Concave models} -- they assume that the total number of faults in software is finite, and that it is possible to achieve fault-free software in finite time. 
\item \textit{S-shaped models} -- they also assume that the total number of faults is finite. and that early testing is not as effective in fault discovery as the testing in the later stages. Therefore, there is a period in which the number of faults is increasing.
\item \textit{Infinite models} -- they assume that it is not possible to develop fault-free software because during fault removal we can introduce new ones.
\end{itemize}

The first step of the SRGM analysis process is to prepare the input data, such as bug reports obtained during testing or operation. The reports should be processed first to identify the faults which caused the failures and only these faults should be considered in SRGM analysis. The bug reports can be further filtered to better fit into the SRGMs assumptions \cite{koz10}.

The second step is to perform a statistical test such as Laplace's trend test \cite{TREND} to determine whether a trend of decreasing number of failures can be observed in the data. If such a trend cannot be observed, it means there is no significant reliability growth in the data and the SRGMs might not provide reasonable results.

Next, the SRGM is selected for the application. It is recommended that multiple models are used and the one describing the data most accurately is selected.
To get results from SRGMs, the model's parameters need to be estimated based on input data. For parameter estimation, either the Maximum likelihood method  or the Minimum least-squares method are typically used \cite{LYU95}. 

After parameter estimation, a goodness-of-fit test (GoF) is performed to evaluate how well the model fits the data to choose the most appropriate SRGM. In related studies (Table \ref{tab:related-sum}), there are several most adopted GoF metrics: the R-squared a.k.a. coefficient of determination (R$^2$) gives the amount of the model explaining the variability of the data~\cite{musa87}, Akaike Information Criterion (AIC) provides the quality of the model~\cite{burn02}, Bayesian Information Criterion (BIC) represents the posterior probability of a model given the data~\cite{burn02}, and  Residual Standard Error (RSE) measuring the standard deviation of the residuals in the regression model~\cite{rse}.

The selected SRGMs can be then used to predict a variety of failure data, such as future failure intensity, number of remaining faults, or testing effort required to achieve a given reliability level.

\section{SRGM \& OSS}
\label{sec:sgrm-oss}

The application of SRGMs in the OSS domain faces several challenges mainly due to underlying assumptions of the traditional SRGMs~\cite{LYU95}. These models were designed with a closed-source software  environment reflecting the waterfall development lifecycle. For instance, the models assume a perfect debugging process in which the detected fault is immediately removed. However, in the OSS context, such assumption would be unrealistic due to the influence of various factors, such as non-homogeneous groups of developers and users both of which can contribute to the bug reports during the development. Additionally, the contributors are equipped with different skills, tools, and resources which may lead to potential regressions with new bugs introduced and thus a rather non-linear debugging process \cite{wang21}.

The main issues in the applicability of SRGMs to OSS are based on the assumptions about the testing and faults repairing process. Some of these assumptions might be violated in the context of modern systems/development practices~\cite{WOOD96}:

\begin{itemize}
    \item \textit{\textbf{Faults are repaired immediately when they are discovered}} -- in reality, faults are not repaired immediately, but this can be partially eliminated by filtering out duplicated reports with a combination of using only resolver reports. 
    \item \textit{\textbf{Fault repairs are perfect}} -- fault repair likely introduces new faults. Because the re-test for the code is not as thorough as the initial testing, the new faults are less likely to be identified.
    \item \textit{\textbf{No new code is introduced during testing}} -- new code is frequently introduced throughout the entire test period (\textit{faults repair} and \textit{new features}), especially for agile development. This is accounted for in parameter estimation since real faults discoveries are used. However, the shape of the curve may be changed (i.e., make it less concave).
    \item \textit{\textbf{Defects are only reported by the product testing group}} -- it does not apply to open source projects, where the whole community of users and developers can submit reports.
    \item \textit{\textbf{Each unit of time is equivalent}} -- it can be accounted for as long as test sequences are reasonably consistent.

    \item \textit{\textbf{Tests represent operational profile}} -- often, the operational profile is uncertain because of the lack of usage statistics. However, nowadays, many applications (\textit{including OSS projects}) collect various usage/telemetry data (\textit{e.g., execution steps/scenarios, frequency of certain inputs, actions}) so this might not be the issue anymore.
    
    \item \textit{\textbf{Failures are independent}} -- it is reasonable except when a section of code has not been as thoroughly tested as other code. 
\end{itemize}

The impact of models' assumption violations is difficult to estimate. For example, introducing a new functionality may make the curve less concave, whereas test re-runs could make it more concave.

Furthermore, in most cases, the testing period is not publicly defined for OSS. Ideally, the reports should be processed first to distinguish the faults which caused the failures, and only these faults should be taken into account in SRGM analysis. This process would be very time-consuming. Therefore, it is possible to process the bug reports to filter out duplicate entries. Afterward, reports are considered as a substitute for a fault \cite{koz10}.

Given the uncertainties about the effects of violating model assumptions, the best strategy for OSS is to try multiple models and see the models that could be better for the prediction of failures.

\section{Related Work}
\label{sec:related-work} 

In the context of OSS, multiple studies focused on a comparison of SRGMs to see which model fits best the defect data originating from OSS repositories. We list these studies in Table \ref{tab:related-sum} detailing the SRGMs used, the best performing models, the number of OSS projects investigated and the GoF measures applied. Mohamed et al. \cite{Mohamed08} focused on the defect classification in OSS projects and how different types of defects impact the SRGM analysis. Rahmani et al. \cite{Rahmain10},  Zhou et al. \cite{zhou05}, Rossi et al. \cite{rossi10}, Fengzhong et al. \cite{Zou08},  and Ullah et al. \cite{Ullah12} studied bug arrival processes in OSS projects in relation to SRGMs. Out of those, authors in \cite{zhou05} and \cite{Ullah12} compared the results also with the industrial and closed-source datasets. Furthermore, the study in \cite{Rahmain10} investigated also the correlation of bug arrival patterns and project popularity with the SRGM results. 

Interestingly, even though the studies overlapped in the applied SRGMs, the results and conclusions were often different. For example, Rossi et al. \cite{rossi10} determined that WE is overall the best performing model. Rahmani et al. \cite{Rahmain10} agreed with the performance of WE in terms of GoF, but in terms of predictive power, they considered WE as the worst model. Zhou et al. \cite{zhou05} concluded that closed-source projects and OSS exhibit similar bug arrival patterns making the traditional SRGMs suitable for OSS datasets with WE being the best model. On the other hand, Fengzhong et al. \cite{Zou08} argued that traditional SRGMs were largely unsuitable and proposed a new type of reliability model instead. Similarly, Wang et al. \cite{wang21} also proposed a new SRGM targeting the OSS projects and showed that their new model outperformed the traditional ones.

\begin{table}[htb]
\footnotesize
\caption{Summary of related studies about the application of SRGM in OSS.}
\label{tab:related-sum}
\begin{tabular}{|p{0.5cm}|p{2.5cm}|p{1.5cm}|p{1cm}|p{1cm}|}
\hline
\textbf{Ref} & \textbf{Models} & \textbf{Best Models} & \textbf{OSS Projects} & \textbf{GoF} \\ \hline
\cite{Rahmain10} & GOS, SCH, WE                      & WE      & 5 & R$^2$       \\ \hline
\cite{Mohamed08} & GO, GOS                           & -{}-    & 2 & R$^2$       \\ \hline
\cite{zhou05}    & WE                                & WE      & 8 & R$^2$       \\ \hline
\cite{rossi10}   & WE, WES, GO, GOM GOS, HD, YE    & WE      & 3 & R$^2$, AIC  \\ \hline
\cite{tamura05}  & GO, HD, LP                        & LP      & 1 & AIC, MSE \\ \hline
\cite{Ullah12}   & MO, HD, GO, GOS, WE, GOM, LOG, YE & GOM, HD & 6 & R$^2$       \\ \hline
\cite{wang21}    & GO, GOS, ISS, PNZ, PZ, WM, L, W                                 & W      & 3 & R$^2$, MSE        \\ \hline \hline
\multicolumn{5}{p{8cm}}{\scriptsize *SCH - Schneidewind model \cite{Rahmain10}, WES - Weibull S-Shaped model \cite{ivanov18}, GOM - Gompertz model \cite{gomperz}, LOG - Logistic model \cite{taghi}, LP - Logarithmic Poisson Execution Time model \cite{tamura05}, MSE - mean squared error, ISS - Inflection S-Shaped model \cite{ohba84}, PNZ - Pham-Nordmann-Zhang model \cite{pham}, PZ - Pham-Zhang model \cite{phamzhang97}, WM - Wang-Mi model \cite{wangmi19}, L - Li model \cite{li11}, W - Wang model \cite{wang21} }
\end{tabular}
\end{table}

\section{Empirical Evaluation}
\label{sec:evaluation}
The goal of the empirical evaluation is to Compare 9 SRGMs on 88 OSS Projects in terms of several GoF measures deriving the performance of the models in the whole dataset, in the single domains, and considering for each project the software releases.

\subsection{Research Questions}

\begin{itemize}
    \item[{RQ1}] \textit{What is the \underline{ranking of the models} based on GoF ?} We investigate the GoF values for all the models applied to OSS projects. We compare mean/SD values for GoF and run Kruskal-Wallis and Dunn's tests to identify the ranking over all the projects.
    
    \item[{RQ2}] \textit{How does the \underline{project's domain} affect the GoF of models?} Projects from different domains might indicate different applicability of SRGMs. We will compare mean/SD values for GoF and run Kruskal-Wallis and Dunn's tests to identify the significant differences of the models.
    
    \item[{RQ3}] \textit{Does project division into \underline{OSS releases} change the applicability of SRGMs?} We compare the applicability of SRGMs to the whole project and to releases of the same project.  We compare mean/SD values for GoF to identify this difference on whole projects and separately on their releases.

    \item[{RQ4}] \textit{How much do different \underline{project attributes} (like lines of code, number of issue reports, number of faults) impact on the applicability of SRGMs? The rationale is to understand if these factors imply different applicability of the models. We segment the projects according to three categories depending on the levels of the attributes and we compare the ranking of the models in terms of R$^2$.
}

\end{itemize}

\subsection{Data Collection}
\begin{figure}
  \begin{center}
    \fbox{\includegraphics[width=0.98\linewidth]{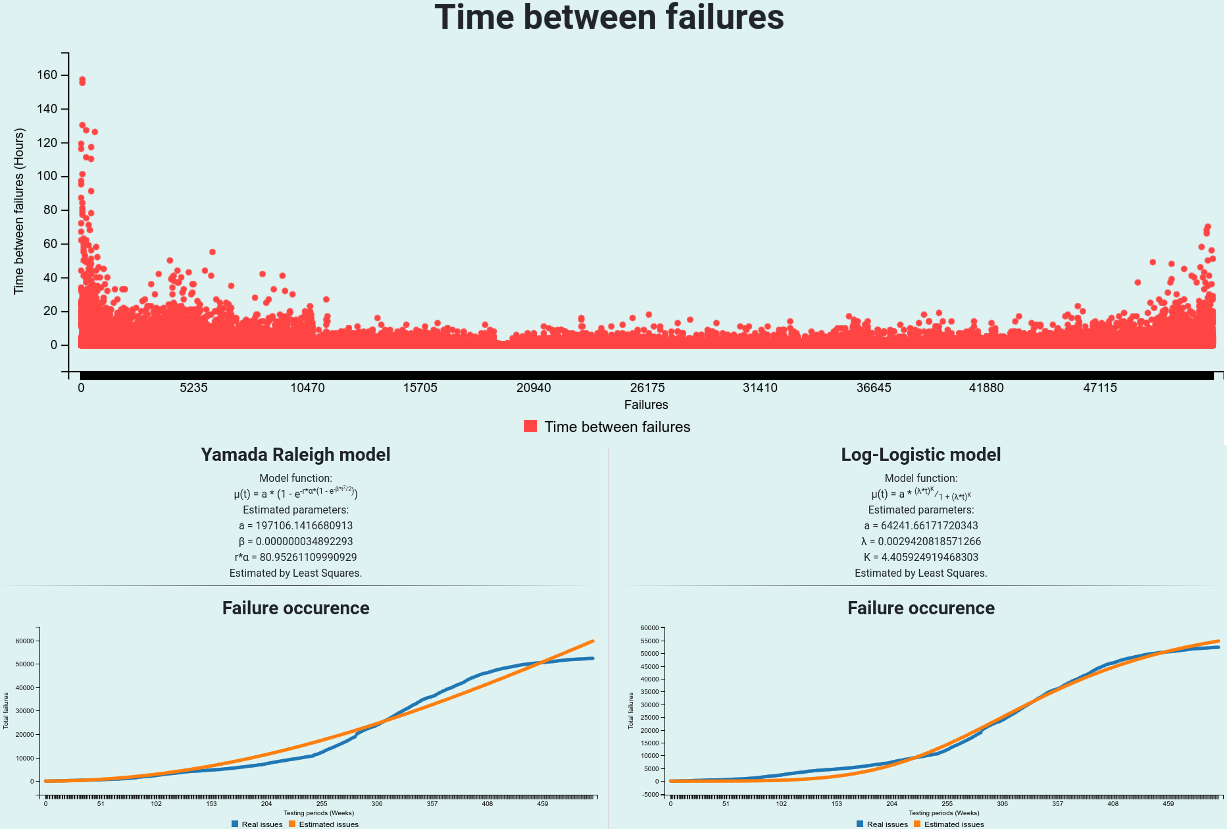}}
  \end{center}
  \caption{Partial STRAIT Output for project Ansible for YR and LL models}
  \label{fig:strait-output}
\end{figure}

We used STRAIT~\cite{chren_Strait} to mine data from GitHub bug tracking repositories. To get a more heterogeneous sample of projects, we have chosen the top ten projects from different topics of the \textit{"Topic Lists"} and combined them with ten more projects from the "Trending List". We focused on the issues that were declared "bug", "error", "fail", "fault" or "defect" excluding any issue that was marked as "duplicated". An example of part of the output of the STRAIT tool is available in Fig. \ref{fig:strait-output}. After filtering the projects which were data repositories, we defined the final set of 88 projects divided into the categories shown in Table \ref{tab:project-categories}. All the mined projects and STRAIT's outputs are available on Figshare~\cite{dataset2022}. More details about the process are also available in~\cite{Micko2022thesis}.

\begin{table}[htb]
\caption{The number of projects based on their categories and the avg. number of defects per project.}
\label{tab:project-categories}
\scriptsize
\setlength\arrayrulewidth{1pt}
\setlength\tabcolsep{4.5pt}
\begin{tabular}{|l|r|p{0.6cm}||l|r|p{0.6cm}|}
\hline
Category                     & \# & Defects (AVG.) & Category           & \# & Defects (AVG.) \\ \hline
\textbf{C1} Admin/monitoring & 14 & 10012          & \textbf{C5} SW Development  & 16 & 1876           \\ \hline
\textbf{C2} Cryptocurrency            & 9  & 2527           & \textbf{C6} System/OS tools & 10 & 8671           \\ \hline
\textbf{C3} DB and data analysis      & 10 & 3327           & \textbf{C7} Text processing & 7  & 833            \\ \hline
\textbf{C4} Multimedia                   & 10 & 5286           & \textbf{C8} Web/Networking  & 12 & 1017           \\ \hline
\end{tabular}
\end{table}

\subsection{Methodology}

To answer the research questions, we fitted the nine SRGM models on the cumulative number of failures for each project. For models to converge, we started with the initial approximation of parameters with 100K algorithm iterations. We then used these initial parameters approximation as input for the original \textit{nls - Nonlinear Least Squares} \cite{nls} method. We allowed a parallel run of parameter estimations in separate threads, so all nine models were estimated at once, reducing the overall time needed for analysis but increasing resource requirements. We run STRAIT \cite{chren_Strait} in a Cloud environment (\texttt{16 threads / 128GB RAM}) for increased performance. Overall, we fitted 792 SRGMs (88 projects x 9 models) for RQ1, RQ2, RQ4 and additionally 261 SRGMs for software releases (29 releases x 9 models) in RQ3.  

To evaluate the Goodness-of-fit (GoF) of the models to the dataset, we used the GoF metrics described in Section \ref{sec:background} (\textit{R$^2$, AIC, BIC, and RSE}).  We then computed mean values and standard deviation ($\sigma$) for GoF metrics for all the projects.

\begin{figure}
  \begin{center}
    \includegraphics[width=0.98\linewidth]{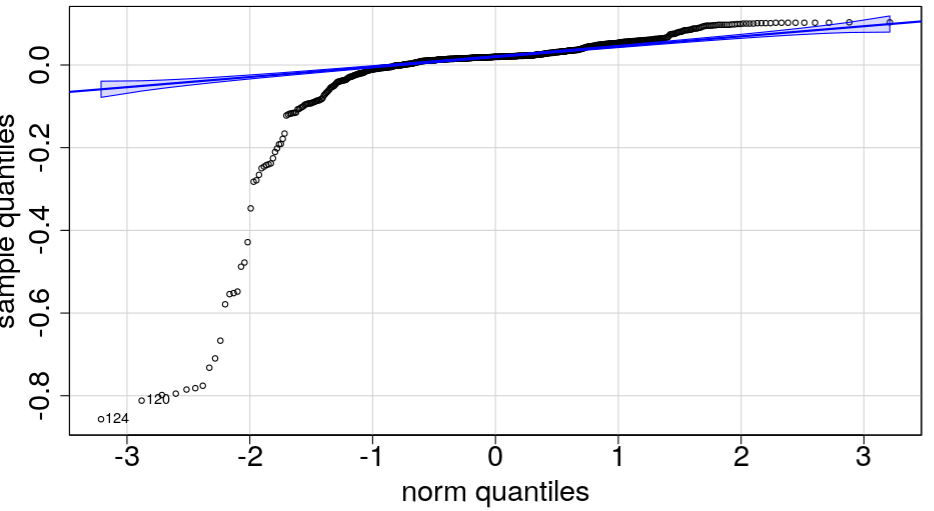}
  \end{center}
  \caption{QQplots of ANOVA's residuals for all models $R^2$}
  \label{fig:qqplots}
\end{figure}

Initially, we planned to use the one-way Analysis of Variance (ANOVA) test~\cite{anova} together with Tukey's Honestly Significant Difference (HSD) test~\cite{tukey} to evaluate the differences between the groups. However, there are several assumptions for the application of ANOVA and HSD (measurements \textit{normally distributed}, \textit{homogeneity of variance} across groups, \textit{independence}) -- in our case we evaluated whether R$^2$ measurements violate strongly any of the assumptions, as ANOVA can be considered robust in case of small violations. After performing Q-Q plots of ANOVA's residuals (Fig. \ref{fig:qqplots}) and Shapiro-Wilk normality test (\textit{W=0.382, p-value $<$ 2.2e-16}, data non-normally distributed) the first assumption on the measurement indicator R$^2$ was strongly violated. For this reason, we opted for Kruskal-Wallis Test with Dunn post-hoc test with multiple testing adjustment based on the Bonferroni correction. The assumptions were not violated for AIC measurements, but we still opted for the more conservative Kruskal-Wallis test also for AIC. For the effect size, we used Eta squared: $\eta^2[H] = (H - k + 1)/(n - k)$, where $H$ is the Kruskal-Wallis test result; $k$ is the number of groups; $n$ is the total number of observations. The rule of thumb for $\eta^2$ is $0.01-<0.06$ (small effect), $0.06-<0.14$ (moderate effect), and $>= 0.14$ (large effect).

\section{Results}
\label{sec:results}
\subsection{RQ1 - Ranking of Models based on GoF}

\begin{figure}[htb]
  \begin{center}
    \includegraphics[width=0.98\linewidth]{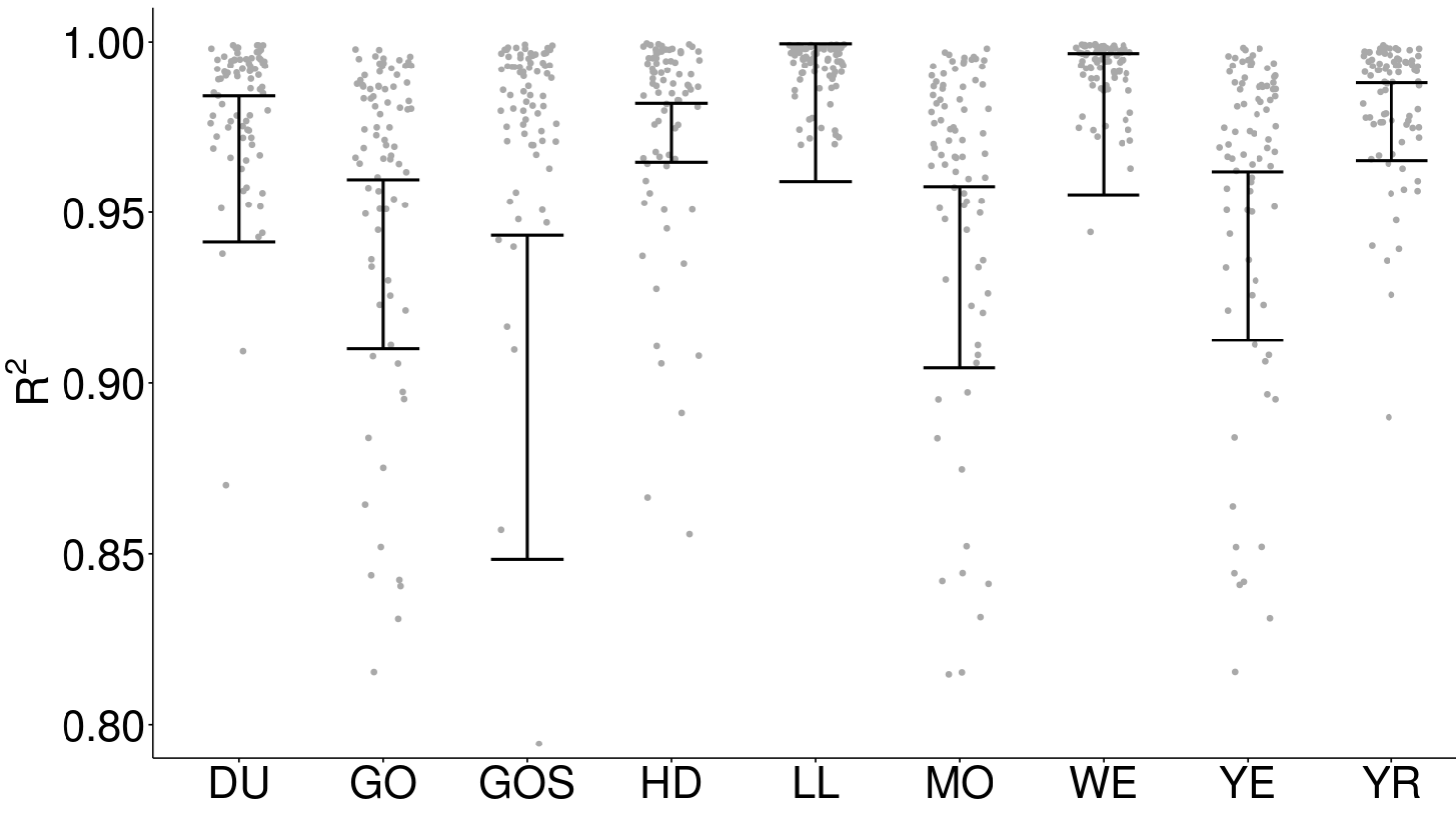}
  \end{center}
  \caption{95\% Confidence Intervals for $R^2$ of the nine models (please note the [1.0-0.8] range - not all datapoints are represented)}
  \label{fig:CI}
\end{figure}

To answer this RQ, we considered 792 SRGMs fitted on the whole dataset with $383\,788$ software defects. Considering all projects' average R$^2$, the Log-logistic model (LL), Yamada Raleigh (YR), Weibull (WE), and Hossain-Dahiya (HD) are the best four models (Table \ref{tab:results-all}). At the same time, Goel-Okumoto S-Shaped (GOS) represents the worse model in terms of R$^2$. However, these considerations are based on the average R$^2$ results from fitting all the models. 
We can look at the distribution of R$^2$ for more insights about the performance of the models. The Kruskal-Wallis rank sum test p-value $\leq$ 0.05 (large effect size $\eta^2[H]=0.191$) indicates that there are statistically significant differences between two or more groups. Overall, we can consider the results of 95\%  R$^2$ Confidence Interval (Fig. \ref{fig:CI}).

\begin{itemize}
    \item GO, GOS, MO, YE have higher variance than other models and they also have worse performance compared to the other models. Dunn post-hoc test shows significantly statistical differences between these and the other models; 
    \item HD and YR are the models that have less variance in terms of R$^2$. Considering the whole dataset, these models reach consistent results, slightly better compared to the best model (LL) considering average R$^2$. Dunn post-hoc test reports significant differences HD-LL (p-value = $0.001$) and YR-LL (p-value = $0.008$) while the difference HD-YR is not significant.
    \item The GOS model is the worse model in terms of average and variance of R$^2$. From the analysis on all the projects, this seems the last model to be suggested for usage.
    \item Since each model has been applied to the same set of projects, we can compare the AIC metric results: the Kruskal-Wallis rank sum test reports that there are non-significant differences among the group of models (p-value = 0.64).
\end{itemize}

\begin{table}[htb]
    \centering
    \caption{RQ1 - GoF of models (ordered by best R$^2$)}
    \scriptsize
    \setlength\arrayrulewidth{0.5pt}
    \begin{tabular}{|p{0.53cm}|c|c|c|c|c|c|}
     \hline
        &\multicolumn{2}{c|}{R$^2$}&\multicolumn{2}{c|}{AIC}&\multicolumn{2}{c|}{RSE}\\\hline
        \textbf{Model} & \textbf{$\mu$} & \textbf{$\sigma$} & \textbf{$\mu$} & \textbf{$\sigma$}& \textbf{$\mu$} & \textbf{$\sigma$}\\
        \hline
        
        LL& 0.979& 0.094 & 3,374& 1771.464 & 83.907&188.828\\\hline
        YR&0.977&0.052&3,806&1847.782&175.468&445.723\\\hline
        WE&0.976&0.096&3,411&1782.277& 81.158& 164.656\\\hline
        HD&0.973&0.039&3,769&1890.365&145.866&380.665\\\hline
        DU&0.963&0.099&3,708&1947.489&147.956&395.543\\\hline
        YE&0.937&0.115&3,889&2087.951&233.414&604.541\\\hline
        GO&0.935&0.115&3,896&2078.743&234.270&604.028\\\hline
        MO&0.931&0.123&3,883&2073.052&230.064&606.065\\\hline
        GOS&0.896&0.220&3,767&1942.052&180.096&438.825\\
        \hline
\end{tabular}
    \label{tab:results-all}
    \end{table}
    
    The first ranking of the Log-logistic (LL) model confirms the findings in \cite{log98} extending the result to other measures of accuracy than the RSE and proving the LL model proposition to capture the increasing/decreasing nature of the failure occurrence rate. Nevertheless, the Weibull  (WE) model also proved its precedence even over a significant sample of projects. However, if we look at the results from the Dunn's test on all 88 projects, the main consideration is that all models apart GOS can be used to fit the cumulative failure data with some degree of confidence.

\begin{mybox}{\textbf{RQ1 Findings}}
Based on 792 fitted SRGMs, considering the R$^2$ metric LL, YR, WE, HD, DU are the best models. GO, GOS, MO, YE show the highest variance  than other models. GOS is in general the worse model in terms of R$^2$.
\end{mybox}

\begin{table*}[htb]

\caption{RQ2 - Overview of the GoF (R$^2$) for the project categories and individual models (top-3 models highlighted).}
\scriptsize
\label{tab:categories-gof}
\setlength\arrayrulewidth{1pt}
\begin{tabular}{|l|cc|cc|cc|cc|cc|cc|cc|cc|}
\hline
 &
  \multicolumn{2}{c|}{\shortstack{\textbf{C1} \\ \tiny $n=126$ \\ \tiny $\leq0.05$ (KWT) \\ \tiny $\eta^2=0.19$}} &
  \multicolumn{2}{c|}{\shortstack{\textbf{C2} \\ \tiny $n=81$ \\ \tiny $\leq0.05$ (KWT) \\ \tiny $\eta^2=0.25$}} &
  \multicolumn{2}{c|}{\shortstack{\textbf{C3} \\ \tiny $n=90$ \\ \tiny $\leq0.05$ (KWT) \\ \tiny $\eta^2=0.10$}} &
  \multicolumn{2}{c|}{\shortstack{\textbf{C4} \\ \tiny $n=90$ \\ \tiny $\leq0.05$ (KWT) \\ \tiny $\eta^2=0.19$}} &
  \multicolumn{2}{c|}{\shortstack{\textbf{C5} \\ \tiny $n=144$ \\ \tiny $\leq0.05$ (KWT) \\ \tiny $\eta^2=0.14$}} &
  \multicolumn{2}{c|}{\shortstack{\textbf{C6} \\ \tiny $n=90$ \\ \tiny $\leq0.05$ (KWT) \\ \tiny $\eta^2=0.19$}} &
  \multicolumn{2}{c|}{\shortstack{\textbf{C7} \\ \tiny $n=63$ \\ \tiny $>0.05$ (KWT) \\ \tiny $\eta^2=0.06$}} &
  \multicolumn{2}{c|}{\shortstack{\textbf{C8} \\ \tiny $n=108$ \\ \tiny $\leq0.05$ (KWT) \\ \tiny $\eta^2=0.13$}} \\ \cline{2-17} 
\textbf{Model} & $\mu$   & $\sigma$ & $\mu$   & $\sigma$ & $\mu$   & $\sigma$ & $\mu$   & $\sigma$ & $\mu$   & $\sigma$ & $\mu$   & $\sigma$ & $\mu$   & $\sigma$ & $\mu$   & $\sigma$ \\ \hline
DU             &\cellcolor{mygreen} 0.989 &\cellcolor{mygreen} 0.008  & 0.982 & 0.016  & 0.850 & 0.269  & 0.976 & 0.027  & \cellcolor{mygreen}0.982 & \cellcolor{mygreen}0.016  & 0.976 & 0.020  & 0.952 & 0.076  & 0.960 & 0.080  \\ \hline
GO             & 0.949 & 0.056  & 0.971 & 0.018  & 0.766 & 0.283  & 0.952 & 0.061  & 0.958 & 0.040  & 0.970 & 0.027  & 0.934 & 0.050  & 0.945 & 0.079  \\ \hline
GOS            & 0.842 & 0.331  & 0.949 & 0.114  & 0.774 & 0.255  & 0.845 & 0.259  & 0.872 & 0.256  & \cellcolor{mygreen}0.987 & \cellcolor{mygreen}0.008  & \cellcolor{mygreen}0.969 & \cellcolor{mygreen}0.026  & 0.980 & 0.017  \\ \hline
HD             & 0.970 & 0.043  & 0.977 & 0.021  & \cellcolor{mygreen}0.988 &\cellcolor{mygreen} 0.011  & 0.968 & 0.072  & 0.966 & 0.033  & 0.982 & 0.028  & 0.954 & 0.041  & \cellcolor{mygreen}0.988 & \cellcolor{mygreen}0.008  \\ \hline
LL             &\cellcolor{mygreen} 0.996 &\cellcolor{mygreen} 0.002  &\cellcolor{mygreen} 0.993 &\cellcolor{mygreen} 0.009  & \cellcolor{mygreen}0.871 & \cellcolor{mygreen}0.277  &\cellcolor{mygreen} 0.993 &\cellcolor{mygreen} 0.007  &\cellcolor{mygreen} 0.989 &\cellcolor{mygreen} 0.009  &\cellcolor{mygreen} 0.994 &\cellcolor{mygreen} 0.009  &\cellcolor{mygreen} 0.984 &\cellcolor{mygreen} 0.009  &\cellcolor{mygreen} 0.994 &\cellcolor{mygreen} 0.055  \\ \hline
MO             & 0.947 & 0.058  & 0.970 & 0.018  & 0.757 & 0.301  & 0.947 & 0.059  & 0.957 & 0.040  & 0.967 & 0.026  & 0.910 & 0.103  & 0.952 & 0.021  \\ \hline
WE             & \cellcolor{mygreen}0.995 & \cellcolor{mygreen}0.003  & \cellcolor{mygreen}0.993 &\cellcolor{mygreen} 0.009  & 0.865 & 0.274  & \cellcolor{mygreen}0.993 & \cellcolor{mygreen}0.008  & \cellcolor{mygreen}0.987 & \cellcolor{mygreen}0.011  &\cellcolor{mygreen} 0.994 &\cellcolor{mygreen} 0.008  & 0.958 & 0.077  &\cellcolor{mygreen} 0.993 &\cellcolor{mygreen} 0.008  \\ \hline
YE             & 0.950 & 0.056  & 0.971 & 0.018  & 0.762 & 0.286  & 0.952 & 0.061  & 0.958 & 0.040  & 0.970 & 0.027  & 0.935 & 0.050  & 0.969 & 0.028  \\ \hline
YR             & \cellcolor{mygreen}0.985 &\cellcolor{mygreen} 0.017  & 0.986 & 0.014  & \cellcolor{mygreen}0.921 & \cellcolor{mygreen}0.165  & \cellcolor{mygreen}0.984 & \cellcolor{mygreen}0.018  & 0.975 & 0.030  & \cellcolor{mygreen}0.987 &\cellcolor{mygreen} 0.009  &\cellcolor{mygreen} 0.970 & \cellcolor{mygreen}0.023  & 0.985 & 0.012  \\ \hline
\end{tabular}
\end{table*}

\subsection{RQ2 - Project Domains}

To answer this RQ, we considered 792 SRGMs fitted on the whole dataset with $383\,788$ software defects and segmented by categories presented in Table~\ref{tab:project-categories}. We analyzed the GoF of the domains C1-C8 by fitting the models on the cumulative failures of the projects in each domain (Table \ref{tab:categories-gof}). All the categories show significant results according to Kruskal-Wallis Test, apart C7 (text processing).   The Log-logistic (LL), the Weibull (WE), and the Yamada Raleigh (YR) are among the top three models in terms of R$^2$ for the majority of the project domains. However, there are some distinctions to put forward: GOS while generally worse in some categories (C1,C3,C4,C5) reaches good performance in other categories (C6,C7,C8), so the application can be considered domain-dependent. The \textit{Database and data analysis (C3)} domain presents interesting results: the concave model Hossain-Dahiya (HD) model fits the best for projects of this domain, followed by LL and WE models, but in general, the results of the majority of the models are the worst compared to other domains, caused by the low failure occurrence rates.
We cannot compare AIC metric results across domains (due to the application to different projects), but when compared in the same domain Kruskal-Wallis test does not report significant differences among the models.


\begin{mybox}{\textbf{RQ2 Findings}}
Divided by domains (C1-C8), the results in terms of R$^2$ metric report some models with a good consistency across domains (e.g., LL, WE, YR). One exception is the database and data analysis domain (C3) in which the lower failure occurrence rate leads to HD model and GO, GOS, MO, YE as the worse models. The GOS model, while statistically worse than all other models when considering the whole dataset, has some domains in which it has low variance and good $R^2$ rankings.
\end{mybox}

\subsection{RQ3 - Project Releases}
To answer this RQ, we used 63 SRGMs (7 projects with releases fitted by 9 models) and 261 SRGMs (29 releases, 9 models, $6\,800$ defects). We looked at the impact of considering project releases compared to projects as a whole when fitting SRGM models to cumulative failure data. Project releases with less than 20 issues indicated a notable bad fit for SRGMs. For such a low number of failure data, it was not feasible to estimate the parameters of the models. For this reason, when analyzing single releases, we omitted project releases with less than 20 faults.

We divided the projects into two groups for the comparison of GoF : \textbf{R}(eleases) -- SRGMs fitted to cumulative failures for projects releases with failure data $>$20, and \textbf{N}(non)\textbf{R}(eleases) -- SRGMs fitted to the cumulative failures for projects as a whole. Considering the \textbf{R} group, the Kruskal-Wallis test was significant (p-value = 4.064e-05, moderate effect size, $\eta2[H] = 0.103$), while for the \textbf{NR} group it was not significant (p-value = 0.437, small effect size $\eta2[H] =0.0006$).

Four models with the highest means of R$^2$ are ranked the same (\textit{LL, WE, DU,  HD}) for both groups (Table \ref{tab:ranking-release}). Using releases or considering the project as a whole does not have an impact on the best models that remain mostly stable (apart YR which gains positions in the ranking). Considering AIC and BIC all the top-3 models are ranked differently when considering software releases. Considering RSE, LL and WE remain the best models in both groups, with the YR model that gains from the last model to the 3rd model in terms of RSE. However, AIC and BIC penalize YR on the release data signalling the model might be overfitting the data (higher R$^2$ and RSE but lower AIC and BIC). 

Looking at Dunn's post-hoc test with Bonferroni adjustment for R$^2$ of the \textbf{R} group, the GOS model is considered to be different from DU, LL, WE models -- as well, YR is statistically different than the LL model. Based on these results, we can deduce that there is not much difference from the models to be selected at the release level, with GOS worse in terms of the performance compared to the other models (Table \ref{hsd5-1}).\\

\begin{mybox}{\textbf{RQ3 Findings}}
Considering projects as a whole or releases mostly does not have an impact on the rankings of the models in terms of R$^2$ (the top-3 models remain the same). Considering AIC and BIC leads to different rankings of models, signalling that some models such as YR might be penalized for overfitting the data. As per RQ1 results, the GOS model is the worse both considering releases and projects as a whole.
\end{mybox}

\begin{table}[htb]
  \centering
\caption{RQ3 - Comparison of rankings of GoF metrics means considering releases (R) or whole projects (NR).}
\label{tab:ranking-release}
\scriptsize
\setlength\arrayrulewidth{1pt}
\begin{tabular}{|c|cc|cc|cc|cc|}
\hline
 & \multicolumn{2}{c|}{R$^2$} & \multicolumn{2}{c|}{AIC} & \multicolumn{2}{c|}{BIC} & \multicolumn{2}{c|}{RSE} \\
 & R & NR & R & NR & R & NR & R & NR \\ \hline
DU & \cellcolor[HTML]{CCFFCC}3 & \cellcolor[HTML]{CCFFCC}3 & 6 & \cellcolor[HTML]{CCFFCC}3 & 7 & \cellcolor[HTML]{CCFFCC}3 & \cellcolor[HTML]{CCFFCC}3 & 5 \\ \hline
GO & 7 & 8 & 4 & 7 & \cellcolor[HTML]{CCFFCC}1 & 7 & 7 & 9 \\ \hline
GOS & 9 & 9 & 9 & 8 & 9 & 8 & 8 & 4 \\ \hline
HD & 4 & 4 & \cellcolor[HTML]{CCFFCC}2 & 5 & \cellcolor[HTML]{CCFFCC}3 & 5 & 4 & 6 \\ \hline
LL & \cellcolor[HTML]{CCFFCC}1 & \cellcolor[HTML]{CCFFCC}1 & 5 & \cellcolor[HTML]{CCFFCC}1 & 6 & \cellcolor[HTML]{CCFFCC}1 & \cellcolor[HTML]{CCFFCC}1 & \cellcolor[HTML]{CCFFCC}1 \\ \hline
MO & 6 & 7 & 8 & 6 & 8 & 6 & 6 & 8 \\ \hline
WE & \cellcolor[HTML]{CCFFCC}2 & \cellcolor[HTML]{CCFFCC}2 & 7 & \cellcolor[HTML]{CCFFCC}2 & 4 & \cellcolor[HTML]{CCFFCC}2 & \cellcolor[HTML]{CCFFCC}2 & \cellcolor[HTML]{CCFFCC}2 \\ \hline
YE & 5 & 6 & \cellcolor[HTML]{CCFFCC}3 & 4 & 5 & 4 & 5 & 7 \\ \hline
YR & 8 & 5 & \cellcolor[HTML]{CCFFCC}1 & 9 & \cellcolor[HTML]{CCFFCC}2 & 9 & 9 & \cellcolor[HTML]{CCFFCC}3 \\ \hline
\end{tabular}
\end{table} 

\begin{table}[htb]
    \centering
    \caption{RQ3 - Result of Dunn's Test (Bonferroni adj) (R$^2$) -- \textit{R} Group} 
    \scriptsize
    \setlength\arrayrulewidth{1pt}
    \begin{tabular}{|c|c|c|c|c|c|}
        \hline
        \textbf{Models} & \textbf{p-value} & \textbf{Models} & \textbf{p-value} & \textbf{Models} & \textbf{p-value}\\
        \hline
        
        GO-DU&1.000&	WE-GO&1.000&	YE-HD&1.000\\\hline
        \cellcolor{mygreen}GOS-DU&\cellcolor{mygreen}0.017&	YE-GO&1.000&	YR-HD&1.000\\\hline
        HD-DU&1.000&	YR-GO&1.000&	MO-LL&1.000\\\hline
        LL-DU&1.000&	HD-GOS&0.293&	WE-LL&1.000\\\hline
        MO-DU&1.000&	\cellcolor{mygreen}LL-GOS&\cellcolor{mygreen}0.000&	YE-LL&1.000\\\hline
        WE-DU&1.000&	MO-GOS&0.115&	\cellcolor{mygreen}YR-LL&\cellcolor{mygreen}0.010\\\hline
        YE-DU&1.000&	\cellcolor{mygreen}WE-GOS&\cellcolor{mygreen}0.000&	WE-MO&1.000\\\hline
        YR-DU&0.538&	YE-GOS&0.064&	YE-MO&1.000\\\hline
        GOS-GO&0.117&	YR-GOS&1.000&	YR-MO&1.000\\\hline
        HD-GO&1.000&	LL-HD&1.000&	YE-WE&1.000\\\hline
        LL-GO&1.000&	MO-HD&1.000&	YR-WE&0.061\\\hline
        MO-GO&1.000&	WE-HD&1.000&	YR-YE&1.000\\\hline
        
    \end{tabular}
    
    \label{hsd5-1}
    \end{table}

\subsection{RQ4 - Project Attributes}
To answer this RQ, we considered all the 88 projects, thus considering 792 SRGMs fitted to $383\,788$ software defects. We segmented the projects according to different attributes:

\begin{itemize}
    \item \textbf{lines of code/size (LOC)} -- the total code size;
    \item \textbf{number of contributors (NOC)} -- the total number of contributors;
    \item \textbf{number of issue reports (NOI)} -- the total number of issue reports created on project;
    \item \textbf{number of faults (NOFA)} -- the total number of faults from issue reports;
\end{itemize}

We divided these attributes into more distinct fragments (\textit{Small (S) / Medium (M) / Large (L)}). The threshold values for these fragments were selected to have at least 10 projects in each segment and to represent a significant subdivision of projects in the three ranges. We derived these thresholds from the indications in previous research on typical domain-specific source code metrics \cite{mori2018} and our evaluation of the distribution of the metrics in the projects considered (Table \ref{tab:categories}).

\begin{center}
\begin{table}[htb]
    \centering
      \caption{Thresholds for category fragments}
    \label{tab:categories}
    \setlength\arrayrulewidth{1pt}
    \begin{tabular}{|l|r|c|r|}
     \hline
        & \textbf{Small} & \textbf{Medium} & \textbf{Large}\\
        \hline
        \textbf{LOC} &  $<$ 10,000 & 10,000 to 100,000 & $>$ 100,000\\\hline
        \textbf{NOC} & $<$ 100 & 100 to 300 & $>$ 300\\\hline
        \textbf{NOI} & $<$ 1,000& 1,000 to 10,000 & $>$ 10,000\\\hline
        \textbf{NOFA} & $<$ 500 & 500 to 5,000 & $>$ 5,000\\
        \hline
    \end{tabular}
    \end{table}
\end{center}

Once the projects were segmented, we looked at the rankings in terms of R$^2$ between the fitted models in the different categories of projects attributes (Table \ref{tab:ranking-attributes}):

\begin{itemize}
    \item In all the projects in which either LOCS, NOC, NOI, NOFA metrics are in the medium and higher ranges, the LL model is usually ranked as the best model in terms of R$^2$. Where the values are instead on the small scale, the YR models is, in the majority of the cases, the best model. This seems to suggest that the two S-Shaped models (LL and YR) have better results in terms of fitting compared to other models;
    \item We found that in most of the cases GOS (the other S-Shaped model) is the worse model when considering the division into categories (S,M,L). Also the MO model reports some consistently lower results in terms of R$^2$.
    \item We compared also the consistency of the rankings of the models for the four categories LOC, NOC, NOI, NOFA in terms of S,M,L division to see the stability of the rankings  in each of the categories. We used the Inter-Rater Agreement (IRA) defined as $IRA = TA / (TR*R) *100$ where \textit{TA~= \#~of agreements in the ratings}, \textit{TR~= \#~of ratings given by each rater}, \textit{R~= \#~of raters} (three in our case, S,M,L). Overall, rankings are more stable in the LOC categories with $IRA=40.7\%$. They are quite low in the NOC categories with $IRR=25.9\%$ and NOI categories with $IRA=22.2\%$. While in the NOFA, there is a very low consistency of the rankings, with only $IRA=3.7\%$.
\end{itemize}

\begin{table}[htb]
  \centering
\caption{RQ4 - Comparison of ranking of the top-2 and bottom-2 models (LOC, NOC, NOI, NOFA) by mean R$^2$ for S, M, L projects.}
\label{tab:ranking-attributes}
\scriptsize
\setlength\arrayrulewidth{1pt}
\begin{tabular}{|c|ccc|ccc|ccc|ccc|}
\hline
 & \multicolumn{3}{c|}{LOC} & \multicolumn{3}{c|}{NOC} & \multicolumn{3}{c|}{NOI} & \multicolumn{3}{c|}{NOFA} \\
 & S & M & L & S & M & L & S & M & L & S & M & L \\ \hline
DU & 5 & 5 & 3 & 6 & 4 & 3 & 5 & 5 & 3 & 5 & 4 & 3 \\ \hline
GO & 6 & 6 & 7 & 5 & \cellcolor[HTML]{f8e2e1}8 & 6 & 6 & \cellcolor[HTML]{f8e2e1}8 & 7 & 6 & \cellcolor[HTML]{f8e2e1}8 & 7 \\ \hline
GOS & \cellcolor[HTML]{f8e2e1}9 & \cellcolor[HTML]{f8e2e1}9 & \cellcolor[HTML]{f8e2e1}9 & \cellcolor[HTML]{f8e2e1}9 & \cellcolor[HTML]{f8e2e1}9 & \cellcolor[HTML]{f8e2e1}9 & \cellcolor[HTML]{f8e2e1}8 & \cellcolor[HTML]{f8e2e1}9 & 6 & \cellcolor[HTML]{f8e2e1}9 & \cellcolor[HTML]{f8e2e1}9 & 5 \\ \hline
HD & \cellcolor[HTML]{CCFFCC}2 & 4 & 4 & \cellcolor[HTML]{CCFFCC}2 & 3 & 5 & \cellcolor[HTML]{CCFFCC}2 & 3 & 5 & \cellcolor[HTML]{CCFFCC}1 & 5 & 6 \\ \hline
LL & 3 & \cellcolor[HTML]{CCFFCC}1 & \cellcolor[HTML]{CCFFCC}1 & 3 & \cellcolor[HTML]{CCFFCC}1 & \cellcolor[HTML]{CCFFCC}1 & 3 & \cellcolor[HTML]{CCFFCC}1 & \cellcolor[HTML]{CCFFCC}1 & 3 & \cellcolor[HTML]{CCFFCC}1 & \cellcolor[HTML]{CCFFCC}2 \\ \hline
MO & \cellcolor[HTML]{f8e2e1}8 & \cellcolor[HTML]{f8e2e1}8 & \cellcolor[HTML]{f8e2e1}8 & \cellcolor[HTML]{f8e2e1}8 & 7 & \cellcolor[HTML]{f8e2e1}8 & \cellcolor[HTML]{f8e2e1}9 & \cellcolor[HTML]{f8e2e1}8 & \cellcolor[HTML]{f8e2e1}9 & \cellcolor[HTML]{f8e2e1}8 & 7 & \cellcolor[HTML]{f8e2e1}9 \\ \hline
WE & 4 & 3 & \cellcolor[HTML]{CCFFCC}2 & 4 & \cellcolor[HTML]{CCFFCC}2 & \cellcolor[HTML]{CCFFCC}2 & 4 & \cellcolor[HTML]{CCFFCC}2 & \cellcolor[HTML]{CCFFCC}2 & 4 & \cellcolor[HTML]{CCFFCC}2 & \cellcolor[HTML]{CCFFCC}1 \\ \hline
YE & 7 & 7 & 6 & 7 & 6 & 7 & 7 & 7 & \cellcolor[HTML]{f8e2e1}8 & 7 & 6 & \cellcolor[HTML]{f8e2e1}8 \\ \hline
YR & \cellcolor[HTML]{CCFFCC}1 & \cellcolor[HTML]{CCFFCC}2 & 5 & \cellcolor[HTML]{CCFFCC}1 & 5 & 4 & \cellcolor[HTML]{CCFFCC}1 & 4 & 4 & \cellcolor[HTML]{CCFFCC}2 & 3 & 4 \\ \hline
\end{tabular}
\end{table} 

\begin{mybox}{\textbf{RQ4 Findings}}
Considering the ranges of LOCs, NOCs, NOIs, and NOFAs, YR and HD models behave the best in the lower range categories, while LL and WE in those with higher ranges. GOS, GO, and MO models are often the worse models in R$^2$ for different categories. NOFA seems to have a high impact on the consistency of the models' rankings.
\end{mybox}

\section{Threats to Validity}
\label{sec:threats}

We only used GoF metrics for the evaluation of the projects. However, one model can have good GoF but bad predictive power~\cite{LYU95}. We plan in further extensions to look into the predictive power of the models for the different domains.
We applied data cleaning and filtering for the issues mined from GitHub repositories, still, some issues might not be related to the system's faults. Furthermore, for fitting the models we used the \textit{minimum least square regression} method to estimate model parameters which is fine for small to medium samples. In contrast, the recommended method for parameter estimation for a significant sample (a large number of faults) is \textit{the maximum likelihood} method. However, for the size of the projects and the comparison of the models, we considered the \textit{minimum least square} method as adequate. Furthermore, there are many project parameters that can potentially affect the results. We do not evaluate them in this study due to space constraints, as the goal of this paper was to be more exploratory than explanatory. However, there were previous studies that looked at such parameters. For example, while the programming language used can lead to certain types of problems, overall, it might not have a significant impact on the number of issue reports. This was the conclusion of an earlier study on 100,000 OSS projects~\cite{bissyande2013popularity}. On the same line, the answers to the RQs are mostly based on the $R^2$ indicator. Even though we report AIC, BIC, and RSE, the stability of the results across the different indicators would require a more in-depth analysis.

\section{Conclusion}
\label{sec:conclusion}
In this paper, we conducted an empirical analysis of SRGMs in the context of the OSS projects. While initial SRGMs were developed without the consideration of agile models, during the years the software development process has evolved towards agile and open source development processes, challenging many of the assumptions of the models that were developed over time.
For this reason, we set up a large empirical study for the application of SRGMs to OSS projects. The main contribution of the study is the evaluation of SRGMs in a large number of projects, considering domains and project releases -- overall, including 88 OSS projects and $\sim$$400K$ software defects, we consider $1\,053$ fitted SRGMs.

Such a large number of fitted SRGMs allowed us to investigate the applicability to OSS on a large scale, looking at the stability of the best models on the whole projects, on releases, on different domains, and according to different projects' attributes.

In general, the results of an extensive number of projects can document the discrepancies in related works about the best models identified. Even the worse model in terms of fitting the cumulative failure data on the whole projects can find good applicability in some domains or specific projects. As such, finding a \textit{one-fits-all} model for OSS, while unrealistic, should further motivate the future investigation of the attributes related to OSS development that might have an impact on the applicability of SRGMs to drive the selection of the most appropriate models. In this sense, while the current paper is exploratory, future works can be in the direction of determining the best predictors from OSS projects that can impact the applicability of SRGMs.

\section*{Acknowledgments}
The work was supported by ERDF "CyberSecurity, CyberCrime and Critical Information Infrastructures Center of Excellence" (No. CZ.02.1.01/0.0/0.0/16\_019/0000822).

Access to the CERIT-SC computing and storage facilities provided by the CERIT-SC Center, provided under the programme "Projects of Large Research, Development, and Innovations Infrastructures" (CERIT Scientific Cloud LM2015085), is greatly appreciated.

\balance



\end{document}